**Title:** A Data-Driven Analysis of the Influence of Care Coordination on Trauma Outcome

**Authors:** You Chen[1] PhD; Mayur B. Patel[2,3,4] MD; Candace D. McNaughton[5] MD, PhD; Bradley A. Malin[1,6,7] PhD

**Author Affiliations:**
[1]Department of Biomedical Informatics, Vanderbilt University Medical Center, Nashville, TN

[2]Department of Surgery, Division of Trauma, Surgical Critical Care, and Emergency General Surgery, Vanderbilt University Medical Center, Nashville, TN

[3]Department of Neurosurgery, Vanderbilt University Medical Center, Nashville, TN

[4]Department of Hearing & Speech Sciences, Vanderbilt University, Nashville, TN

[5]Department of Emergency Medicine, Vanderbilt University Medical Center, Nashville, TN

[6]Department of Biostatistics, School of Medicine, Vanderbilt University, Nashville, TN

[7]Department of Electrical Engineering & Computer Science, School of Engineering, Vanderbilt University, Nashville, TN

**To Whom Correspondence Should be Addressed:**

You Chen, Ph.D.
2525 West End Ave, Suite 1475
Department of Biomedical Informatics
Vanderbilt University
Nashville, TN 37203 USA
Email: you.chen@vanderbilt.edu

# KEY POINTS

QUESTION: **What are the differences in length of stay (LOS) for trauma inpatients with similar age, illness, procedural burden, and insurance type, but were managed by different care coordination patterns?**

FINDINGS: **In this observational study of 5,588 adult patients (age > 18) hospitalized for trauma at a single center between 2013 and 2015, the hospital LOS for the care coordination pattern with the highest degree of collaboration was 14 hours shorter than other care coordination patterns.**

MEANING: **In trauma care, promoting highly collaborative care coordination may reduce LOS.**

# ABSTRACT

**IMPORTANCE:** Care coordination can improve efficiency and reduce health care expenditures, particularly in trauma settings, where a high-throughput of admissions necessitates communication and collaboration between a wide array of healthcare professionals across a range of operational areas in the organization. However, at the same time, the complexity of such a setting makes it challenging to investigate the relationship between coordination plan and patient outcome.

**OBJECTIVE:** To test the hypothesis that variation in care coordination is related to LOS.

**DESIGN** We applied a spectral co-clustering methodology to simultaneously infer groups of patients and care coordination patterns, in the form of interaction networks of health care professionals, from electronic medical record (EMR) utilization data. The care coordination pattern for each patient group was represented by standard social network characteristics and its relationship with hospital LOS was assessed via a negative binomial regression with a 95% confidence interval.

**SETTING AND PATIENTS** This study focuses on 5,588 adult patients hospitalized for trauma at the Vanderbilt University Medical Center. The EMRs were accessed by healthcare professionals from 179 operational areas during 158,467 operational actions.

**MAIN OUTCOME MEASURES:** Hospital LOS for trauma inpatients, as an indicator of care coordination efficiency.

**RESULTS:** Three general types of care coordination patterns were discovered, each of which was affiliated with a specific patient group. The first patient group exhibited the shortest hospital LOS and was managed by a care coordination pattern that involved the smallest number of operational areas (102 areas, as opposed to 125 and 138 for the other patient groups), but exhibited the largest number of collaborations between operational areas (e.g., an average of 27.1 connections per operational area compared to 22.5 and 23.3 for the other two groups). The hospital LOS for the second and third patient groups was 14 hours ($P = 0.024$) and 10 hours ($P = 0.042$) longer than the first patient group, respectively.

**CONCLUSIONS AND RELEVANCE:** The care coordination pattern with the largest degree of collaboration (e.g., highest connectivity and density) exhibited the shortest LOS. We believe these findings serve as evidence for opportunities in codifying or refining coordination patterns for trauma patients.

Trauma is the most common cause of death in the Western world [1-2] due, in part, to the wide variability in the types of injuries and complications to which the patients are subject. As a result, care coordination in trauma often involves interactions between a multidisciplinary group of healthcare professionals (e.g., anesthesiologists, surgeons, emergency room physicians, respiratory therapists, nurse practitioners, radiographers, neurosurgeons, and various types of nurses) who are distributed across time and space [3]. This type of care coordination has been shown to generally reduce the time required for resuscitation [3], improve the overall survival rate of trauma patients [3-4], and improve the efficiency and cost effectiveness of care [3-5].

Given the potential for better outcomes, there is great interest in optimizing the care coordination process (e.g., composition of a trauma team and communication between team members). Towards this goal, various auditing (e.g., video review, observer review and medical notes review) [3, 6], simulation (e.g., simulators who educate team members on communication, cooperation and leadership) [7] and data-driven programs (e.g., tools to visualize care team structures) [21-23] have been proposed to assess and refine care coordination processes. Auditing and simulation programs have generally adapted a top-down strategy to design or refine expert-based care coordination routines, which results in a heavy reliance on the individual healthcare professional and the defined coordination plans are insufficient to satisfy dynamic changes in healthcare environments [38-41]. Thus, data-driven programs have been proposed to infer dynamic care team structures and communication patterns between team members involved in a care coordination process [11-12, 22-23]. However such programs are limited in that they do not investigate the relationship between these patterns and patient outcome, such as length of stay (LOS) [8-9].

Electronic medical record (EMR) systems can bridge this gap, and in many cases they already capture information sharing, coordination, and documentation longitudinally, thus providing a large quantity of operational activities from a diverse collection of healthcare professionals [13-14, 22, 26-29]. This type of data has shown promise for inferring healthcare organizational patterns [11-12] and analyzing patient outcomes [37]. Thus, in this study we leverage EMR data to infer trauma-centered coordination patterns and quantify their relationship with hospital LOS.

**Methods**

**Study Materials**

This study focuses on 5,588 patients who were assigned to a single center's trauma service (Vanderbilt University Medical Center) and completed an inpatient stay between 2013 and 2015. During this period, 5,547 employees, affiliated with 179 operational areas (e.g., mental health center, neuro intensive care unit, and neurosurgery clinic) in the medical center, accessed the EMRs of these patients. This entailed 158,467 unique actions (e.g., uploading a clinical note or medication administration). These access actions were relied upon to infer care coordination patterns (e.g., networks of healthcare employees).

Our study leverages a single-institution EMR system as a special case to explore trauma-centered coordination patterns and quantify their relationship with hospital LOS. This medical center provides state- and national-verified Level 1 trauma care for a geographic region spanning 65,000 square miles and handles over 3,000 acute trauma admissions annually [10]. Additionally, information associated with a patients' hospitalization, as well as employees' utilization of such information, is documented in a homegrown EMR system that has been central to clinical activities since the 1990s [11-13].

To assess the relationships between care coordination patterns with patient outcomes, we extracted hospital LOS for each patient, where LOS is measured as the duration between admission and discharge of a patient encounter. The mean and media of length of stay was 158 and 160 hours respectively. We excluded patients who died during their trauma service to focus on coordination patterns indicative of completion in hospital care. Given that LOS may also be related with additional confounding factors (e.g., patient age, degree of illness, procedure burden and insurance type), we extracted and tested the influence such factors for each patient as well.

The EMRs for the patients in this study contained 3,612 distinct International Classification of Diseases, Ninth Version (ICD-9) billing codes, 1,627 distinct Current Procedural Terminology (CPT) codes and 8 insurance programs (e.g., Medicaid, Medicare Part A). The minimum, mean and maximum patient age was 18, 47 and 100 respectively.

We acknowledge that ICD-9 codes are insufficient to represent accurate patient illness. To mitigate bias due to practices in insurance billing, we rely on the phenome-wide association study (PheWAS) vocabulary, which was introduced to group ICD-9 codes together and reduce variability in the definitions of clinical concepts in secondary data use scenarios [16-17]. Upon translating each ICD-9 code, the data consisted of 1,010 PheWAS codes.

**Study Design**

In preparation for this study, we composed a set of cohorts by defining cases (e.g., patients with longer LOS) and controls (e.g., patients with shorter LOS). Then, care coordination patterns (e.g., networks of healthcare employees) were inferred for cases and controls, respectively, and their relationship with hospital LOS was examined.

Given such a design, it would seem natural to investigate the influence of care coordination patterns on differences in LOS. However, care coordination patterns may not be designed or established according to the criteria (e.g. LOS) we leveraged to define the cases and controls. Thus, the care coordination patterns inferred from such defined cases and controls may be segments of real coordination and could not represent the exact networks of healthcare employees.

As a result, we construct cohorts by grouping patients according to inferred coordination patterns. This was accomplished by i) inferring the care coordination patterns and then ii) applying the inferred patterns to compose the cohorts. To do so, patients in each built group will share similar care coordination patterns. In this case, it is appropriate to control for care coordination variations in each patient group and, thus, ascertain if such variations are related with LOS.

Our study design consists of three components: i) learn patient groups according to care coordination patterns; ii) quantify care coordination patterns by standard social network characteristics; and iii) assess the relationships between coordination patterns and hospital LOS.

*Grouping Patients By Care Coordination Patterns*

We use a binary matrix $A$ to represent the commitment of healthcare employees' actions to EMRs. Specifically, the value of a cell $A(i,j)$ is 1 if a healthcare worker $i$ committed an action to the EMR of patient $j$ and 0 otherwise. We will leverage information in $A$ to derive care coordination patterns and patient groups. However, it has been shown that care coordination patterns inferred at a level of operational areas are more stable and interpretable than patterns learned at the level of healthcare employees [11,18]. Thus, we transform $A$ into a new matrix $A'$, where each cell stores the number of actions that all healthcare employees from a specific operational area committed to a patient's EMR.

Since, a patient record is usually worked on by a small subset of the healthcare employees, $A'$ is a sparse matrix. We apply a spectral co-clustering model to $A'$ to uncover groups of patients according to their coordination patterns. We rely on this method because it has been shown to be robust in high-dimensional sparse matrices [2-3]. Patients who share similar care coordination patterns were grouped via a spectral co-clustering model [15, 24-25]. This methodology employs matrix decomposition techniques, and formalizes co-clustering as a bipartite graph partitioning problem [2-3]. The details of the patient grouping process are in **Supplement S1**.

*Quantifying the Characteristics of Care Coordination Patterns*

To ascertain if variations in care coordination patterns were associated with hospital LOS, we represent each care coordination pattern as a network of operational areas and then quantify the networks via social network characteristics. For each group of patients, we infer a coordination pattern, in a network form, to show how healthcare employees coming from these operational areas interacted with each other to provide care for the group of patients. Each node in a network corresponds to an operational area and each edge weight of two operational areas is the cosine similarity of interactions on patients' EMR between healthcare professionals from these two operational areas [30]. The details for the edge weighting process are in **Supplement S1**. We define a care coordination pattern as the network of operational areas affiliated with a patient group.

We leverage standard social network characteristics to quantify each care coordination pattern. Specifically, these characteristics correspond to average node degree, average weighted node degree, graph density, clustering coefficient and average path length [31, 34]. The definitions of these characteristics are defined as:

- **Average node degree:** Calculated by summing the degree of for each node (i.e., the number of edges connected to it) and dividing by the total number of nodes.
- **Average weighted node degree:** Calculated by summing the weighted degree of a node (i.e., the sum of weights of edges connected to it) and dividing by the total number of nodes.
- **Graph density:** The ratio of the number of edges observed to the number of possible edges.

- **Cluster coefficient:** The average clustering coefficient for all nodes. The cluster coefficient of a node is the ratio of existing edges connecting a node's neighbors to each other to the maximum possible number of such edges. A high clustering coefficient for a network is another indication of a small world phenomena.
- **Average path length**: Calculated by summing shortest path lengths between all pairs of nodes and dividing by the total number of pairs. This indicates the number of steps, on average, it takes to move from one node of the network to another.

Subnetworks may exist within each care coordination pattern, so we further infer communities of healthcare employees. This is accomplished through a heuristic algorithm that optimizes the modularity of a network [33]. A high modularity indicates a dense connectivity of operational areas within communities and a sparse connectivity between communities.

*Assessing the Relationship between Care Coordination Patterns and LOS*

We apply a generalized linear regression model with negative binomial distributions [19] to test the influence of care coordination patterns on the differences in LOS for each pair of patient groups. The negative binomial distribution has been shown to achieve the best performance to transform hospital LOS, whose distribution does not follow a normal distribution [32]. We applied an analysis of variance (ANOVA [43]) with a 95% confidence interval to test the significance of differences in LOS for pairs of groups.

We further investigated if the differences in LOS were correlated with the potential confounding factors, which we incorporated in the regression models. This was accomplished by testing for differences in the distributions of PheWAS codes, procedural codes, insurance programs and ages between each pair of patient groups. Specifically, for each pair of patient groups, we compare the similarity in the distributions of the aforementioned factors through a Pearson correlation coefficient (PCC) [20]. This similarity score is in the range (-1,1), where a 1 indicates a positive direct correlation, 0 indicates no correlation, and a -1 indicates a negative direct correlation. If each pair of patient groups exhibits a high PCC with a predefined significance (>95% confidence level) for each factor, then we consider them to be sufficiently similar. If a pair of groups exhibit similar distributions in terms of these factors, then their corresponding care coordination patterns likely handle a similar patient population. This would suggest that the different care coordination patterns manage similar patients, but with different hospital LOS. Further details about this assessment are in **Supplement S2**.

## Results

### Patient Groups and Care Coordination Patterns

The co-clustering approach discovered three patient groups, which we refer to as $P_1$, $P_2$, and $P_3$. These groups were composed of 428, 1,353, and 3,807 inpatients, respectively. Additionally, the strategy discovered three operational groups, which we refer to as $O_1$, $O_2$, and $O_3$. The relationship between the patient and operational groups is depicted as a heatmap in Fig 1(a). In this figure, each point indicates the number of actions that healthcare workers from a certain operational area committed to a patient's EMR. Given that there are 179 operational areas, we defer the descriptions for these area to **Supplement S3**. For each patient group, it can be seen that all three operational groups are involved, but with different care coordination patterns (as shown in Fig 1(b), 1(c) and 1(d)).

The care coordination patterns for $P_1$, $P_2$, and $P_3$, are shown in Fig 1(b), 1(c) and 1(d), respectively. Each coordination pattern is made up of communities of operational areas which were represented with different colors (e.g., blue, green, red and purple). The width of an edge between two operational areas indicates the strength of collaboration between their employees, while the size of a node is proportional to the degree (connections) of that node. We rely on connections to emphasize the importance of an operational area within a care coordination pattern.

### Quantified Care Coordination Patterns

The quantities of network characteristics for care coordination patterns are reported in Table 1. It can be seen that the coordination pattern for patient group $P_1$ has the smallest number of operational areas (102), but the highest amount of collaboration (e.g., 27.14 average degree, 7.02 average weighted degree, 0.27 graph density, 0.77 cluster coefficient, and 1.43 average path length) between employees from these operational areas. There are several notable findings to report here. First, the coordination pattern for $P_1$ has the highest average degree (in comparison to coordination patterns for $P_2$ and $P_3$), which indicates that healthcare employees in the care coordination pattern of $P_1$ have more connections with other team members than employees in the other two patterns. Second, the coordination pattern of $P_1$ has the highest graph density and cluster coefficient, which demonstrates that there are more interactions and collaborations between team members than the other patterns. Third, the coordination pattern of $P_1$ has the shortest average path length, which shows a pair of team members in the coordination pattern tend to have a more direct line of communication than in the other two patterns.

**Relationship between Coordination Patterns and LOS**

The relationships between care coordination patterns and LOS are depicted in the top of Table 2. Table 2 shows the LOS of $P_1$ is the shortest, being 14 hours shorter than $P_2$ and 10 hours shorter than $P_3$. The differences in LOS between $P_1$ and $P_2$, as well as $P_1$ and $P_3$, were found to be significant at the 95% confidence level. There were no significant differences in LOS between $P_2$ and $P_3$. This may to be due to the fact that the differences in the characteristics of care coordination patterns (e.g., graph density (0.17 vs. 0.18) and cluster coefficient (0.70 vs. 0.73)) are not that large enough to support the discovery of a small difference in LOS.

The differences between the patient groups in terms of PheWAS codes, procedure codes, insurance programs and ages are depicted in the bottom of Table 2. It can be seen that each patient group exhibits similar distributions in terms of i) PheWAS codes (>0.97 PCC, $p < 2.62 \times 10^{-26}$); ii) procedure codes (>0.98 PCC, $p < 1.67 \times 10^{-26}$); iii) insurance programs (>0.98, $p < 1.73 \times 10^{-5}$); and iv) ages (>0.88, $p < 3.36 \times 10^{-12}$). These results suggest that the three care coordination patterns provided care for the similar patients, but lead to different LOS. This indicates an opportunity to shorten LOS via a refinement of high collaborative care coordination pattern.

**Discussion**

To the best of our knowledge, this is the first study to use EMR data to study the relationship between care coordination patterns and inpatient hospital LOS among trauma patients. In doing so, it fills a gap in knowledge about the relationship between care coordination (patterns) and health outcome (LOS). The findings show that the shortest LOS is associated with care coordination patterns with high connectivity, collaborative capability and low communication cost.

There are several limitations to this study that we wish to highlight to guide future research in this area. First, the findings suggest that more care coordination leads to better health outcomes; however, the reason behind this finding is unclear. Specifically, our investigation focused on the statistical analysis and not the semantics of why collaboration occurs. For instance, it is unclear why, the three care coordination patterns lead to different LOS if they provide care for such similar patients populations. Although we found no significant difference between patient groups in terms of PheWAS codes, procedure codes, insurance types and age, additional factors (e.g., specific traumatic injury) should be investigated.

Second, there may be discordance between the care coordination actions that manifest in the physical clinical world and those documented in the EMR system. Although the criteria of meaningful use for EMR systems has been in existence for a number of years (and is now its third stage), there are still coordination actions in the physical clinical world that are not documented in the EMR systems [42]. This missing information may influence the care coordination patterns and patient groups we inferred from the EMR data.

Third, apart from LOS, other types of care quality measurements (e.g., survival rates, days in the intensive care unit and mortality) need to be included to assess the performance of care coordination. LOS is a critical characteristic to measure health expenditures, for instance, a shorter LOS for several major procedures (e.g., abdominal hysterectomy), do not seem to be associated with any adverse outcomes and result in modest financial saving to the health service [35-36]. However, relations between LOS and care quality are unclear [35-36]. Thus, it is a necessity to quantify influence of care coordination patterns on a combination of LOS and care quality measurements (e.g., readmission rate). In this case, there is an opportunity to derive care coordination patterns with small health expenditure but high care quality.

Fourth, this investigation was based on data from a single academic medical center. Replication of this study using data from other healthcare organizations is necessary to confirm these findings.

**Conclusions**

This study leveraged data-driven methodologies (e.g., spectral co-clustering and network analysis) and statistical models (e.g., regression models and person correlation coefficient) to analyze EMR data to infer care coordination patterns and quantify their influences on LOS for trauma patients. This study showed that care coordination patterns with the high level of collaboration is associated with a shorter hospital LOS for trauma patients. This finding is notable because it suggests there is an opportunity to mitigate lengthy stays via the selection of highly collaborative coordination processes throughout a hospital or healthcare system.

**Online Supplements**

S1: Spectral Co-Clustering Algorithm to Infer Groups of Patients and Operational Areas

S2: Measuring the Distributions of Potentially Confounding

S3: A List of Names for Operational Areas

**Funding**

This research was supported, in part, by the National Institutes of Health under grants R00LM011933 and R01LM010685.

**Competing Interests Statement**

The authors have no competing interests to declare.

**Contributors**

YC performed the data collection and analysis, methods design, hypotheses design, experiments design, evaluation and interpretation of the experiments, and writing of the manuscript. CM and MP performed the hypotheses design, interpretation of experiments and writing of the manuscript. BM performed hypotheses design, evaluation and interpretation of the experiments, and writing of the manuscript.

**Figures**

**Figure 1.** a) A heatmap of the inferred patient groups $P_1$ (428 patients), $P_2$ (1,353), and $P_3$ (3,807) and operational area groups $O_1$ (27 areas), $O_2$ (86 areas) and $O_3$ (66 areas). b-d) Three care coordination patterns for $P_1$ - $P_3$. Each pattern is composed of operational areas coming from all three operational area groups, but with different patterns of interaction (e.g., network structures) among the employees.

## Tables

**Table 1** A comparison of care coordination patterns in terms of their size, average degree, average weighted degree, graph density, average cluster coefficient and average path length.

| Coordination Pattern Metric | Patient Group | | |
|---|---|---|---|
| | $P_1$ | $P_2$ | $P_3$ |
| Number of Operational Areas | 102 | 138 | 125 |
| Degree Average | 27.14 | 23.38 | 22.47 |
| Weighted Degree Average | 7.02 | 5.78 | 5.31 |
| Graph Density | 0.27 | 0.17 | 0.18 |
| Cluster Coefficient Average | 0.77 | 0.70 | 0.73 |
| Path Length Average | 1.43 | 2.16 | 1.52 |

**Table 2.** The difference in LOS for each pair of patient groups are reported, along with the p-value.

| | Influences of Care Coordination Patterns on Length of Stay | |
|---|---|---|
| | LOS Difference (Hours) | P Value |
| $P_1 - P_2$ | -14 | 0.024* |
| $P_1 - P_3$ | -10 | 0.042* |
| $P_3 - P_2$ | -4 | 0.104 |

| Patient Groups | Similarity analysis of factors potentially confounding the relationship between LOS and care coordination patterns. | | | | | | | |
|---|---|---|---|---|---|---|---|---|
| | PheWAS codes | | Procedure codes | | Insurance programs | | Age | |
| | PCC | P value | PCC | P value | PCC | P value | PCC | P value |
| $P_1$ v. $P_2$ | 0.9791 | $2.62 \times 10^{-26}$ | 0.9855 | $1.67 \times 10^{-26}$ | 0.9808 | $1.73 \times 10^{-5}$ | 0.8858 | $3.36 \times 10^{-12}$ |
| $P_1$ v. $P_3$ | 0.9866 | $1.31 \times 10^{-27}$ | 0.9934 | $2.56 \times 10^{-29}$ | 0.9938 | $5.98 \times 10^{-7}$ | 0.9479 | $1.86 \times 10^{-17}$ |
| $P_2$ v. $P_3$ | 0.9929 | $4.32 \times 10^{-29}$ | 0.9929 | $3.87 \times 10^{-29}$ | 0.9810 | $1.68 \times 10^{-5}$ | 0.9644 | $4.82 \times 10^{-20}$ |